\documentclass{aa} 

\usepackage{natbib}
\usepackage{hyperref}
\usepackage[T1]{fontenc}
\usepackage{ae,aecompl}
\usepackage{rotating}
\usepackage{color}
\usepackage{graphicx}

\usepackage{txfonts}

\begin{document}

 \title{Unveiling the nature of donor stars of ULXs in NGC 1559}

  \author{S. Allak\inst{1} and A. Akyuz\inst{1,2}}

   \institute{Space Science and Solar Energy Research and Application Center (UZAYMER), University of Çukurova, 01330, Adana, T\"urkiye\\
       \email{0417allaksinan@gmail.com}
     \and
       Department of Physics, University of Çukurova, 01330, Adana, Türkiye\\
       \email{aakyuz@cu.edu.tr}
       \thanks{}
       }

  \date{Received ---, ----; accepted ----, ----}

 
 \abstract
{X-ray data provide insights into accretion processes and the compact objects of ultraluminous X-ray sources (ULXs), while optical and infrared (IR) observations help identify the donor star and surrounding environment. Together, these approaches shed light on the origins of the high X-ray luminosities observed in ULXs}
{This study examines the optical and infrared properties of eight ULXs in NGC 1559 using archival data from the \textit{Hubble Space Telescope} (\textit{HST}) and \textit{James Webb Space Telescope} (\textit{JWST}).
To constrain the nature of the donor stars of the ULXs, photometric results were obtained from the temporal, spectral energy distributions (SEDs), and color-magnitude diagrams (CMDs). Furthermore, the long-term and spectral characteristics of only a ULX X-1 were investigated.}
{ULX counterparts were determined from astrometric calculations. The long-term light curves and SEDs were constructed to interpret the origin of the optical and IR emissions. The masses and ages of donor star candidates were determined using CMDs. To constrain the mechanism of X-ray emission, the time-averaged spectrum and long-term light curve of the X-1 were obtained.}
{Unique optical and IR counterparts for ULXs X-14 and X-24 were determined, while only optical counterparts were detected for X-1 and X-18. Our findings indicate that the optical emission originates from the donor stars of X-14 and X-24, whereas for X-1 and X-18, it is dominated by the accretion disk. 
In addition, the X-1 exhibits long-term X-ray variability spanning orders of magnitude.}

  {}


  \maketitle
%

\section{Introduction}

Ultraluminous X-ray sources (ULXs) are a distinct subclass of X-ray binaries (XRBs) characterized by an isotropic X-ray luminosity (L$_{X}$) exceeding 10$^{39}$ erg s$^{-1}$, which surpasses the Eddington limit for a 10 M$_{\odot}$ BH. There is a growing consensus that ULXs are predominantly powered by supercritical accretion onto neutron stars (NSs) or stellar mass BHs, as discussed in recent reviews \citep{2017ARA&A..55..303K,2021AstBu..76....6F,2023NewAR..9601672K,2023arXiv230200006P}. However, some studies propose the existence of intermediate-mass BHs (IMBHs) within the range of 10$^{2}$ to 10$^{4}$ M$_{\odot}$, accreting at sub-Eddington rates \citep{2007Ap&SS.311..203R,2012MNRAS.423.1154S,2013MNRAS.436.3262C,2013ApJ...774L..16P}. In the case of supercritical accretion, geometric beaming and/or super-Eddington accretion onto stellar-mass compact objects could explain their high luminosities. \citep{2009MNRAS.393L..41K,2015MNRAS.454.3134M,2018ApJ...857L...3W}. Temporal variability is a key characteristic of ULXs, with quasi-periodic oscillations (QPOs) and coherent pulsations being observed, providing insights into the nature of the compact object \cite{2015MNRAS.446.3926A,2014Natur.514..202B,2018MNRAS.476L..45C,2020ApJ...895...60R}. Furthermore, the detection of cyclotron resonance scattering features in the X-ray spectrum of a ULX suggests the presence of NS candidate \citep{2018NatAs...2..312B,2022ApJ...933L...3K}.

Moreover, identifying optical candidates and analyzing their properties to constrain the age, mass, and possible spectral type of the donor star makes important contributions to ULX studies. The optical emission observed in ULXs can arise from the accretion disk, the donor star, or a combination of both, though many studies suggest that this emission is often dominated by reprocessed radiation from an irradiated accretion disk \citep{2011ApJ...737...81T,2012ApJ...750..152S,2014MNRAS.444.2415S,2018MNRAS.480.4918A,2019ApJ...884L...3Y,2022MNRAS.510.4355A,2024MNRAS.527.2599A}. Long-term optical light curves have revealed sinusoidal modulations, indicating super-orbital or orbital periods for some optical counterparts \citep{2009ApJ...690L..39L,2022MNRAS.510.4355A,2022MNRAS.517.3495A}.
Identifying donor stars in ULXs is challenging due to the faintness of their apparent magnitudes and their locations in crowded, star-forming regions, necessitating high spatial resolution detectors like \textit{JWST} \citep{2024MNRAS.527.2599A}. Some ULX counterparts, particularly those bright in the near-infrared (NIR) band, have been suggested to be red supergiants (RSGs), though recent studies indicate that many NIR counterparts may not be sufficiently red to be RSGs, possibly due to unresolved sources in distant galaxies \citep{2014MNRAS.442.1054H,2023MNRAS.526.5765A}. Additionally, ULXs without detectable optical counterparts in previous studies may still have counterparts in \textit{JWST} images, possibly hidden by hot dust or circumbinary disks \citep{2020ApJ...890..150C,2024MNRAS.527.2599A}.

This study aims to identify the IR and optical counterparts of the eight ULXs identified by \cite{2023ApJ...956...41M} in the NGC 1559 galaxy. NGC 1559 is a barred spiral galaxy located at a distance of 12.6 Mpc, as determined using the Tully–Fisher relation \citep{2013AJ....146...86T}. Although a wide range of distance values (9-23 Mpc) have been determined for this galaxy using different methods in the NED \footnote{(NASA/IPAC Extragalactic Database)}, we have adopted the value of 12.6 Mpc for comparison in the X-ray analysis with \cite{2023ApJ...956...41M}. The galaxy is also notable for hosting four supernovae in the last 40 years, including SN 1986L which has been the primary focus of \textit{Chandra} observation. The galaxy, classified as SB(s)cd, features fragmented spiral arms with high star formation rates but likely lacks an active galactic nucleus (AGN) \citep{1991rc3..book.....D}. In this work, we seek to investigate the potential donor stars of ULXs by utilizing data from {\it JWST} and {\it HST}, with an emphasis on their masses, ages, and emission mechanisms within these systems. Additionally, we conducted a detailed study of the long-term variability and energy spectra of ULXs in NGC 1559 using {\it Chandra} and {\it Swift/XRT}. The paper is organized as follows: Section \ref{sec:2} presents the properties of the galaxy NGC 1559, along with the target ULXs, and provides details of the observations. Data reduction and analysis of X-ray and multi-wavelength observations are discussed in Section \ref{sec:3}. Section \ref{sec:4} presents results and discusses the properties of the ULXs. Finally, Section \ref{sec:5} summarizes the main findings of this study.

\section{Target sources and observations} \label{sec:2}

\subsection{Ultra-luminous X-ray Sources in NGC 1559}

\begin{figure}
 \resizebox{\hsize}{!}{\includegraphics{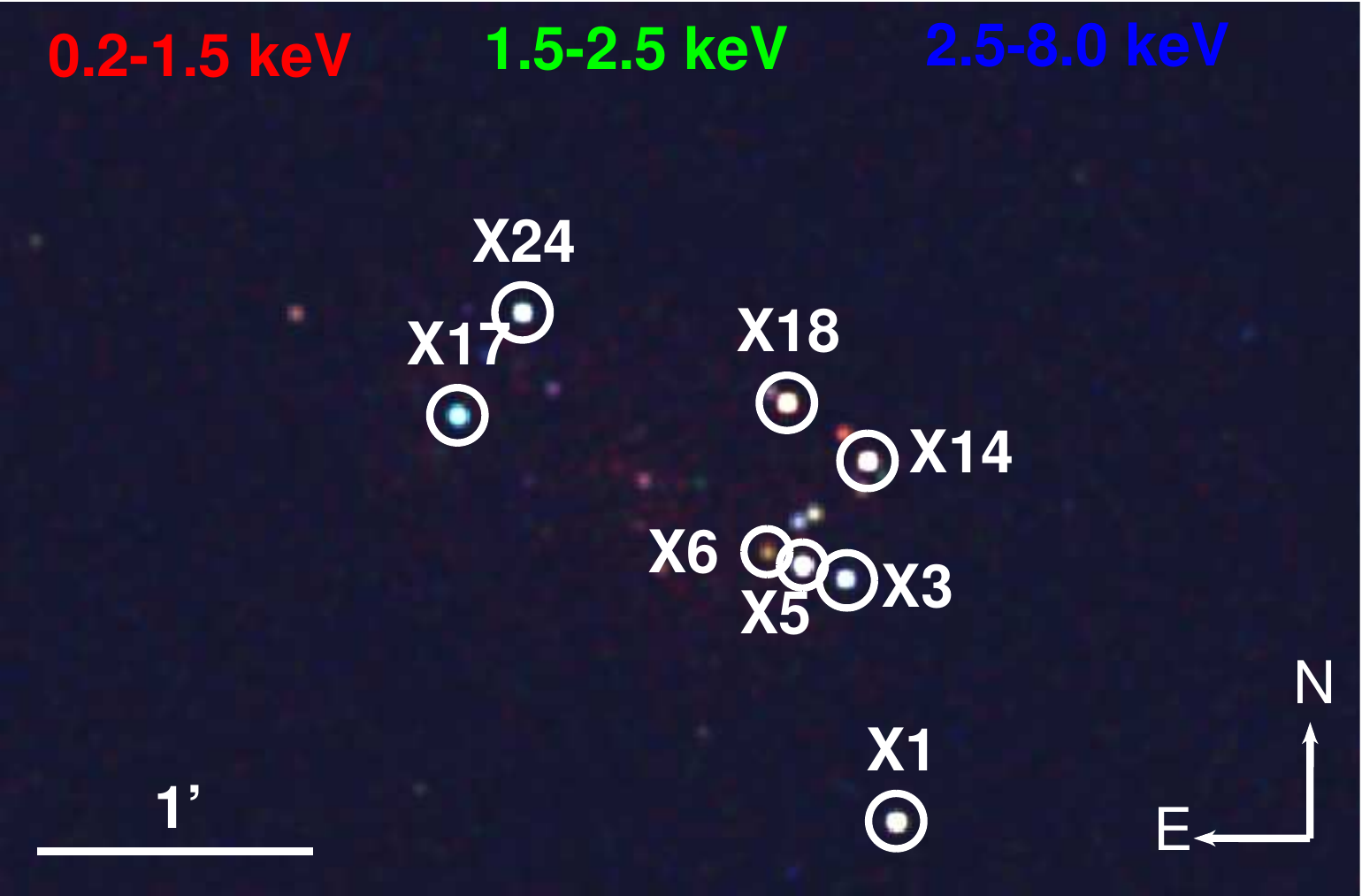}}
 \caption{A false-color \textit{Chandra} image of the galaxy NGC 1559. The energy ranges used for the color image are highlighted. The image has been smoothed using a 5 arcsecond Gaussian, and the ULXs are indicated with white circles.}
 \label{F: pos_x}
\end{figure}

The primary target sources of this study are the eight ULXs (X-1, X-3, X-5, X-6, X14, X-17, X-18, X-24) displayed in Figure \ref{F: pos_x}. In \citep{2023ApJ...956...41M}, ULXs were defined by the criterion L$_{X}$ > 10$^{39}$ erg s$^{-1}$, which we adopted in this study. These sources exhibit X-ray luminosities ranging from a minimum of
10$^{39}$ erg s$^{-1}$ to of a maximum 7.98 $\times$ 10$^{39}$ erg s$^{-1}$ in the (0.3-7) keV energy band and, along with various spectral properties, as detailed in their work. Below, we summarize their characteristics.

X-3, X-5, X-14, and X-17 all display hard spectra with photon indices ($\Gamma$) between 1.47 and 1.85, indicative of non-thermal emission consistent with the presence of a compact object such as a BH or NS.
Among these sources, X-3 has the lowest
L$_{X}$ value at $\sim 2 \times 10^{39}$ erg s$^{-1}$, while X-17 has the highest value at $\sim 8 \times 10^{39}$ erg s$^{-1}$. In contrast, X-6 and X-18 present softer spectra with $\Gamma$ = 3.16 and 2.34, respectively, suggesting the presence of a thermal component or distinct accretion mechanisms. X-6, with a luminosity of $\sim 1 \times 10^{39}$ erg s$^{-1}$, stands out due to its particularly soft spectrum, while X-18, at $\sim 3.6 \times 10^{39}$ erg s$^{-1}$, may represent a different emission mechanism linked to its accretion environment.

One of the remaining two ULXs, X-1, has an L$_{X}$ of $\sim$ 4 $\times$ 10$^{39}$ erg s$^{-1}$ and $\Gamma$ is 1.8, indicating a hard X-ray spectrum. The source also shows long-term variability on a timescale of 10 to 100 days. The isolated location of X-1 within NGC 1559 reduces the risk of contamination from nearby sources. The other ULX,X-24, is particularly noteworthy due to its L$_{X}$ of $\sim$ 3.7 $\times$ 10$^{39}$ erg s$^{-1}$ and the detection of a periodicity around 7500 s. This periodicity is indicative of an orbital period within a compact binary system, suggesting that X-24 may host a stellar-mass BH or NS. The quasi-sinusoidal variation observed in the light curve strengthens the hypothesis that X-24 represents a unique case of a ULX within NGC 1559. In this study, X-1 and X-24 are given priority because of their distinct characteristics. X-1 displays high luminosity and variability, while X-24 exhibits a notable periodicity.

\begin{figure}
 \resizebox{\hsize}{!}{\includegraphics{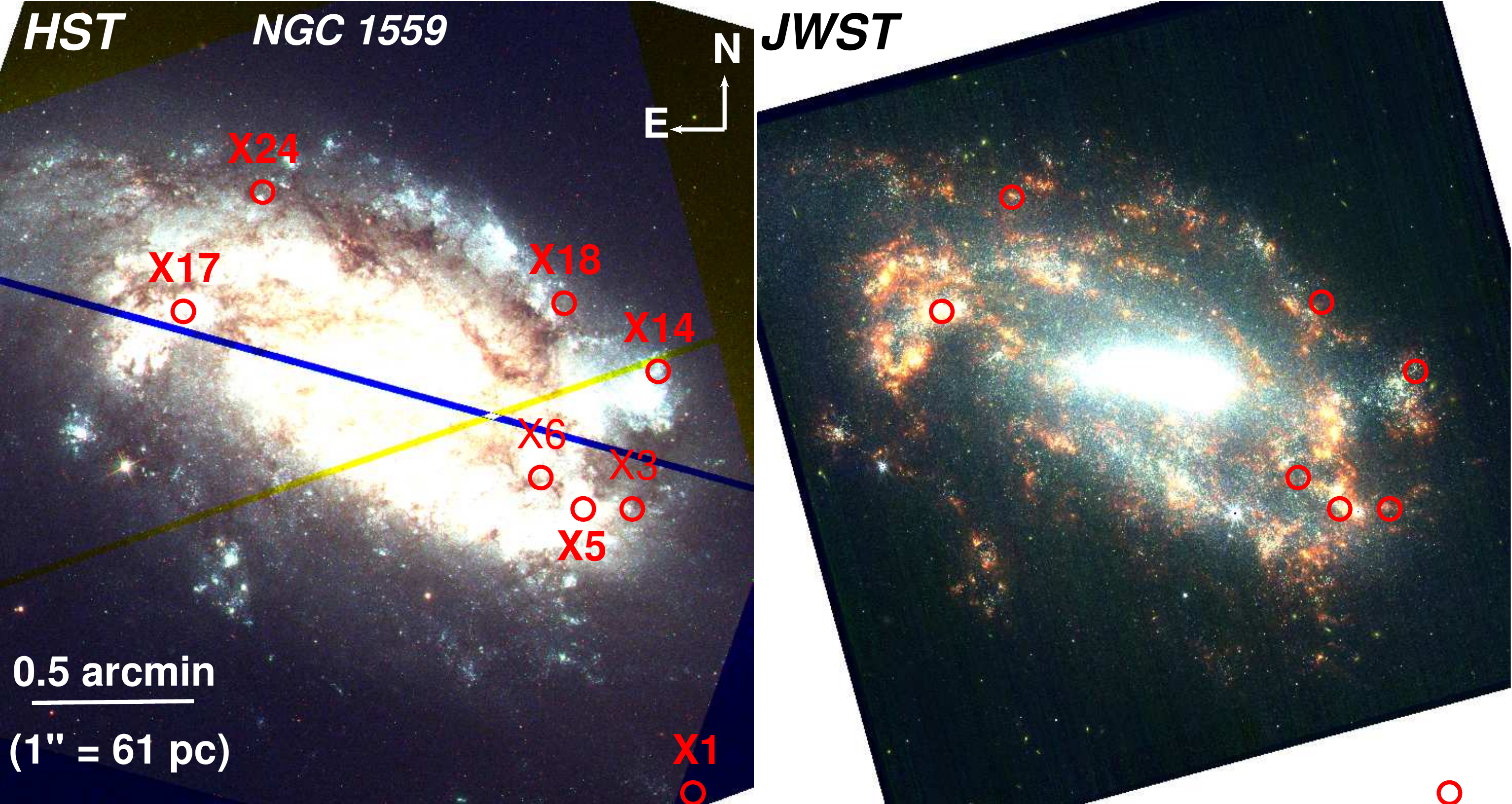}}
 \caption{The RGB (red:green:blue) images of NGC 1559 from {\it HST} (left) and {\it JWST} (right). The filters used for the \textit{HST} RGB image are F814W, F555W, and F438W, and for \textit{JWST}, they are F355M, F300W, and F275W, respectively. The ULXs are marked with red circles on both images. Note that in the \textit{JWST} image, the X-1 source is outside the area.}
 \label{F:rgb}
\end{figure}

\subsection{X-ray and multi-wavelength observations of NGC 1559}

NGC 1559 was observed by {\it Swift/XRT} a total of 56 times between 2005 and 2016 (target IDs: 30252, 34132, 35896, 92169), with exposure times ranging from 25.1 s to 5302.3 s. Additionally, the galaxy was observed for 46 ks by \textit{Chandra ACIS} (\textit{Advanced CCD Imaging Spectrometer}) in 2016 (Obs ID: 16745). These observations provide comprehensive data for studying the X-ray sources in NGC 1559, with a particular focus on the potential variability of ULXs over time. Besides the \textit{Chandra} and {\it Swift/XRT} observations, NGC 1559 was also observed by \textit{XMM-Newton} with a 14.4 ks exposure. However, this observation was excluded from the current study, as no new findings were obtained beyond those reported in the study by \cite{2023ApJ...956...41M}.

The NGC 1559 was observed by {\it HST} employing the \textit{WFC3/UVIS1} (\textit{The Wide Field Camera 3}) in 2015, 2017 and 2019. Moreover, the {\it GAIA}\footnote{https://www.cosmos.esa.int/web/gaia/dr3} source catalog is used for astrometry calculations in this study. The {\it JWST} observed the spiral galaxy NGC 1559 from June to September 2023. The galaxy was imaged using the \textit{JWST/NIRCam} with filters F090W, F150W, F277W, F300M, and F335M, as well as the \textit{JWST/MIRI} with F770W and F2100W filters. We noted that the ULXs in this galaxy are located in highly crowded regions, where sources that are distinguishable in \textit{NIRCam} images often blend in \textit{JWST/MIRI} images.
 The details of all the\textit{HST/WFC3/UVIS1} observations are given in Table \ref{T:obs}. RGB images of the galaxy from {\it HST} and {\it JWST} are displayed in Figure \ref{F:rgb}. Moreover, between 2017 and 2021, NGC 1559 was observed 13 times using the {\it HST/WFC3/IR} F160W filter (Proposal IDs: 15145, 16250). These observations have enabled us to investigate long-term IR variations for counterparts.

\begin{table*}
\centering
\caption{The log of \textit{HST} optical and \textit{JWST} infrared observations}
\begin{tabular}{cccccccc}
\hline
Observatory/Instrument & Filter & Date & Exp (s) & Proposal ID\\
\hline
 JWST/NIRCAM/IMAGE & F090W &2023-06-30 &3349.9 &1685\\
 JWST/NIRCAM/IMAGE &F150W &2023-06-30 &4208.8 &1685\\
 JWST/NIRCAM/IMAGE &F277W &2023-06-30 &7558.7 &1685\\
 JWST/NIRCAM/IMAGE &F090W &2023-07-15 &1674.9 &1685\\
 JWST/NIRCAM/IMAGE &F150W &2023-07-15 &2104.4 &1685\\
 JWST/NIRCAM/IMAGE &F277W &2023-07-15 &3779.3 &1685\\
 JWST/NIRCAM/IMAGE &F300M &2023-09-14 &214.7 &3707\\
 JWST/NIRCAM/IMAGE &F335M &2023-09-14 &386.5 &3707\\
\hline
HST/WFC3/UVIS1&F336W &2015-10-30 &1080 &14253\\
HST/WFC3/UVIS1&F438W &2015-10-30 &1215 &14253\\
HST/WFC3/UVIS1&F606W &2015-10-30 &1350 &15145\\
HST/WFC3/UVIS1&F814W &2017-09-12 &930 &15145\\
HST/WFC3/UVIS1&F555W &2017-09-15 &930 &15145\\
HST/WFC3/UVIS1&F814W &2017-09-26 &930 &15145\\
HST/WFC3/UVIS1&F555W &2017-10-02 &930 &15145\\
HST/WFC3/UVIS1&F814W &2017-11-07 &930 &15145\\
HST/WFC3/UVIS1&F275W &2019-04-06 &4666 &15654\\
\hline
\end{tabular}
\label{T:obs}
\end{table*}

\section{Data Reduction and Analysis} \label{sec:3}

\subsection{Source Detection \& Photometry}

For all drizzled images \textit{HST/WFC3/UVIS1} of this galaxy, point-like sources were detected with the {\it daofind} task, and aperture photometry of these sources was performed using the {\it DAOPHOT} package \cite{1987PASP...99..191S} in {\scshape IRAF}\footnote{https://iraf-community.github.io/} (Image Reduction and Analysis Facility). To perform aperture photometry, 3 pixels (0$\arcsec$.15) aperture radius and for the background, nine pixels were chosen. The Vega magnitudes were derived from instrumental magnitudes using zero point magnitudes (ZPM) taken from the \textit{WFC3/UVIS1} and IR was taken from the study \cite{2022AJ....164...32C}. To derive the aperture corrections a similar approach of \cite{2023MNRAS.526.5765A} was followed. Moreover, to determine the positions and photometry of the sources, the photometry tools provided by the {\scshape photutils v1.82} package\footnote{https://photutils.readthedocs.io/en/stable/index.html\#} Astropy was utilized. For the {\it JWST} data, the process included background estimation, source detection, and photometric analysis, following the methodology outlined in \cite{2023MNRAS.526.5765A}. Given the significant variation in background levels across the images, the {\scshape photutils} background task was applied to estimate local background levels. Each detected source was identified as being more than 3-$\sigma$ above the local background. Aperture photometry was then conducted using a circular aperture with a 3-pixel radius, with the background contribution subtracted based on an annulus positioned nine pixels away from the source center. 

\subsection{Astrometry and determination of counterparts} \label{X-5}

Precise astrometry is necessary to determine both the NIR and optical counterparts of the eight ULXs, whose positions are shown in Figure \ref{F: pos_x}. Optical counterparts of ULXs can be investigated using {\it HST} and {\it Chandra} observations, owing to their excellent spatial resolution. The reference sources were searched by comparing {\it Chandra} image with {\it HST} observations. For this, 32 X-ray sources were detected in {\it Chandra} ACIS-S image by using {\it wavdetect} tool in {\it Chandra} Interactive Analysis of Observations ({\scshape ciao})\footnote{https://cxc.cfa.harvard.edu/ciao/}. Wavelets of 2, 4, 8, and 16 pixels were used, along with a detection threshold of 10$^{-6}$. Only 27 out of 32 X-ray sources matched the {\it HST} images. The 27 X-ray sources were compared with optical point sources, and only one reference source which is the ULX X-5 was found for astrometric calculations. Therefore, {\it GAIA/DR3}\footnote{https://www.cosmos.esa.int/web/gaia/data-release-3} optical source catalog was compared with the \textit{Chandra} X-ray sources since the single source is relatively less reliable for astrometric calculations. Taking account of the same shift directions, including the ULX X-5, three reference sources were found between both {\it Chandra}-{\it GAIA}. The astrometric offsets between {\it Chandra} and {\it GAIA} were found as -0$\arcsec$.33 $\pm$ 0$\arcsec$.17 for R.A and -0$\arcsec$.27 $\pm$ 0$\arcsec$.01 for Decl. with 1-$\sigma$ errors.

\begin{figure}
 \resizebox{\hsize}{!}{\includegraphics{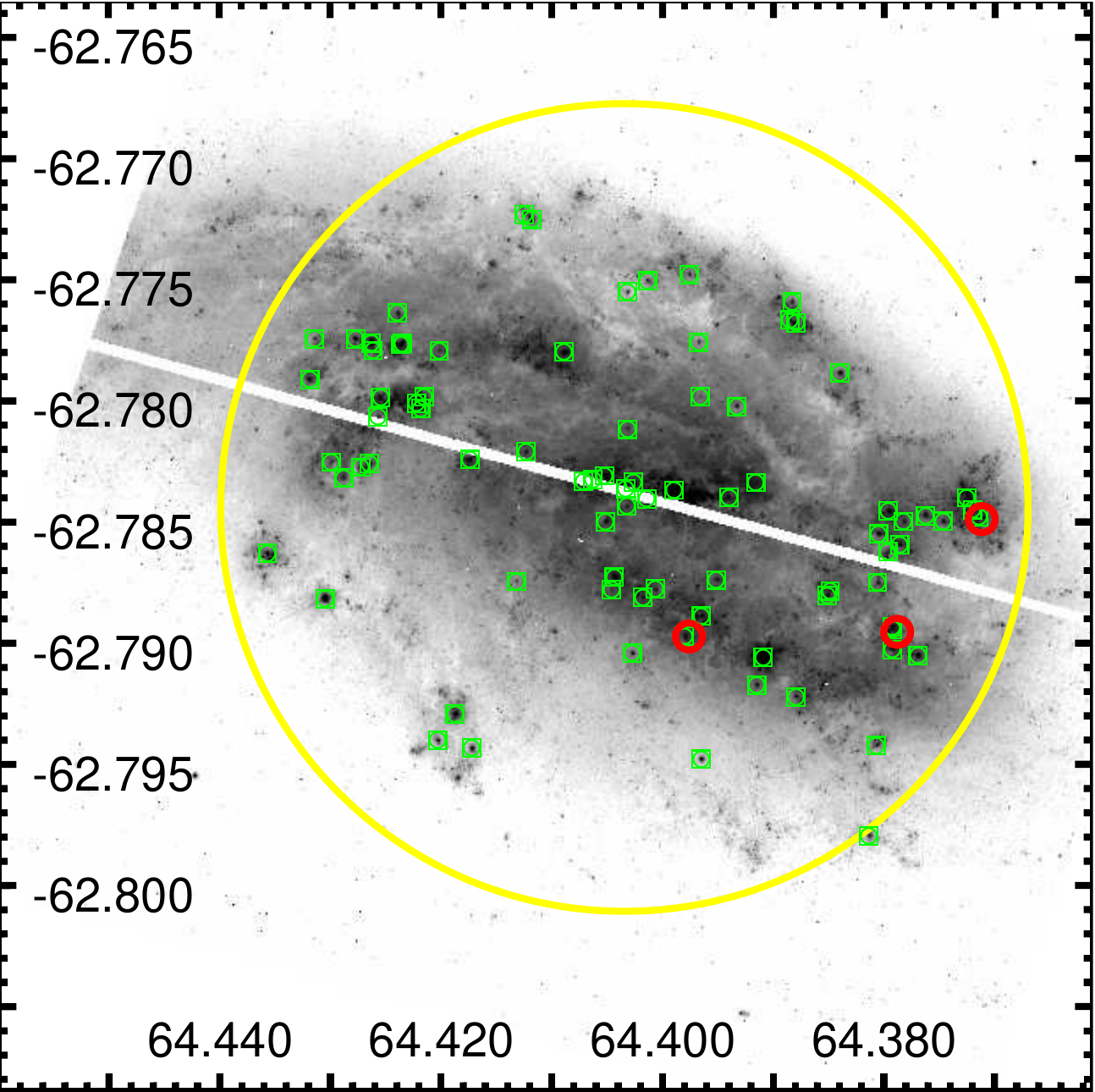}}
 \caption{Positions of {\it Chandra} (red circle) and {\it GAIA (green box)} reference sources used in astrometric calculations are shown on the \textit{HST} image. The solid yellow circle indicates the area within which reference sources are searched, with a radius of 1 arcmin. Here the y and x axes indicate declination (Decl.) and right ascension (R.A), respectively.}
 \label{F:astrometri}
\end{figure}

Furthermore, to find the counterparts of the ULXs, \textit{GAIA} sources were compared with the sources identified in the {\it HST} images, and at least 8 references with matching positions were identified. Positions of all the references used for astrometric calculations are displayed on the {\it HST} F555W image in Figure \ref{F:astrometri}. Astrometric offsets between {\it GAIA} and {\it HST} were found as -0$\arcsec$.006 $\pm$ 0$\arcsec$.001 for R.A and -0$\arcsec$.010 $\pm$ 0$\arcsec$.001 for Decl. with 1-$\sigma$ errors. Following our study of \cite{2022MNRAS.517.3495A}, the positional error radius was derived as 0.38 arcsec at a 90 percent confidence level. This radius corresponds to 23.33 pc based on the adopted distance of NGC 1559 (1 arcsec $\simeq$ 61 pc; \cite{2013AJ....146...86T}). Moreover, relative astrometry was performed between {\it NIRCam} and {\it HST} images to determine the NIR counterparts using {\it GAIA/DR3} source catalog. After adequate reference sources were identified, offsets between {\it JWST/NIRCam} - {\it HST/UVIS1} were derived as -0$\arcsec$.023 and 0$\arcsec$.023, respectively at 1-$\sigma$ confidence level. Additionally, corrections were made by performing relative astrometry both between the filters of \textit{HST} and within the filters of \textit{JWST}. Finally, the X-ray positions were corrected on the basis of all these shifts. The \textit{Chandra} coordinates and corrected coordinates of the ULXs are given in Table \ref{T:summary}.

\begin{table}
\caption{The \textit{Chandra} and corrected X-ray coordinates of the ULXs. Note: Only the optical coordinates of X-1 are based on the F606W image, while the other sources are based on the F555W image.}
\begin{tabular}{cccccccccllll}
\hline
ULXs & RA $\&$ Decl. & RA $\&$ Decl. \\
 &  (X-ray)  & (Corrected)\\
\hline
X-1 &  64.366594 -62.804978 & 64.366863 -62.804894 \\
X-3 &  64.373321 -62.790356 & 64.373586 -62.790259    \\
X-5 &  64.378959 -62.789540 & 64.379224 -62.789453     \\
X-6 &  64.383585 -62.788780 & 64.383850 -62.788683    \\
X-14 & 64.370329 -62.783264 & 64.370595 -62.783168   \\
X-17 & 64.424393 -62.780507 & 64.424658 -62.780410   \\
X-18 & 64.381031 -62.779763 & 64.381295 -62.779666   \\
X-24 & 64.415849 -62.774313 & 64.416114 -62.774216   \\
\hline
\label{T:summary}
\end{tabular}
\end{table}

\begin{figure}
 \resizebox{\hsize}{!}{\includegraphics{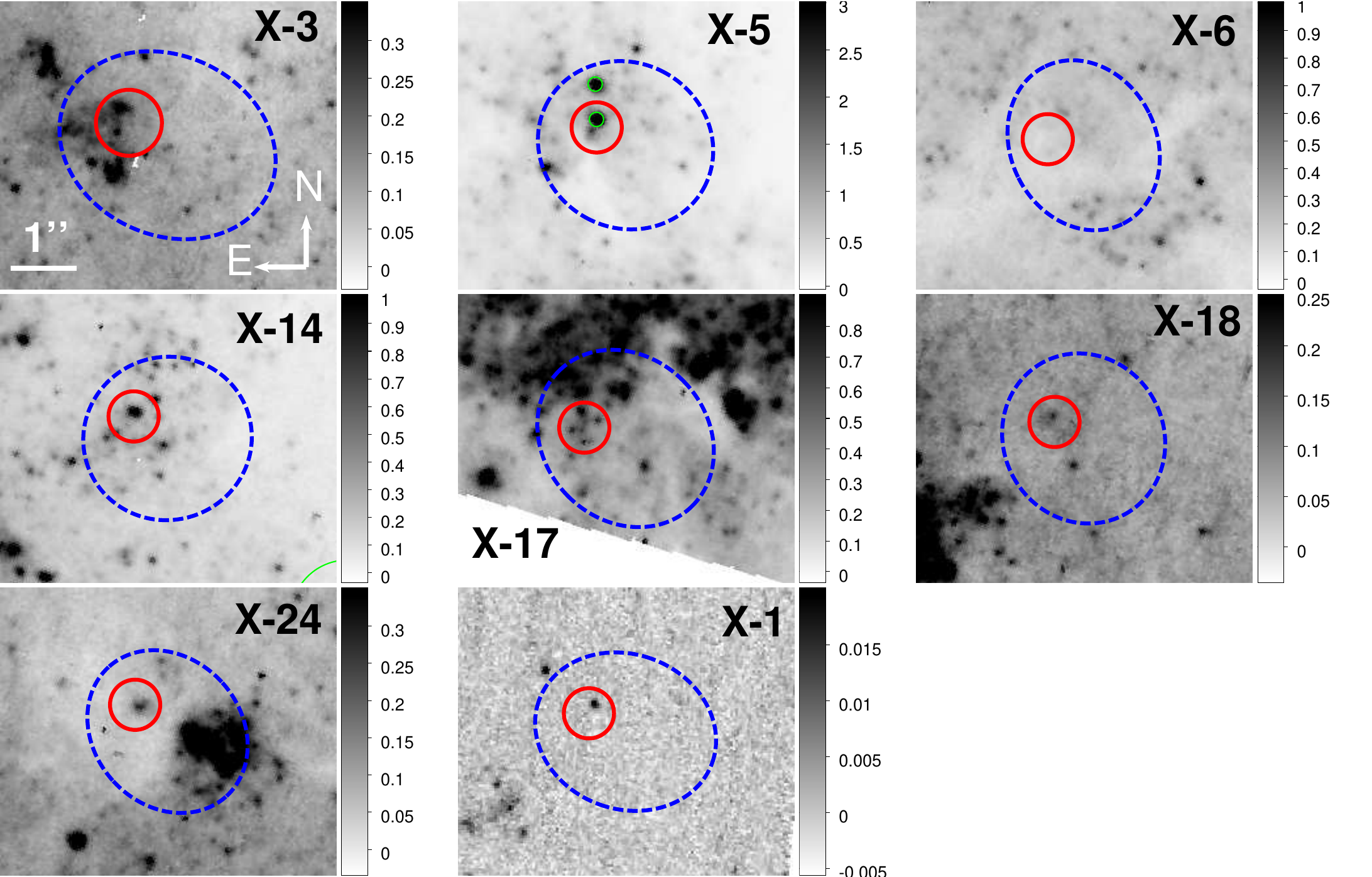}}
 \caption{ The {\it HST} F555W image of corrected X-ray positions (solid red circles) and also \textit{Chandra} source detection error ellipses (dashed blue ellipses) of the ULXs. Since X-1 is not observed in \textit{HST} F555W filter, its position is shown on the F438W image. The numbers shown for the color bars are in units of electrons s$^{-1}$. 
 Solid green circles within the error radii of X-5 indicate \textit{GAIA} sources. All panels are the same scale, with north up and east to the left.}
 \label{F:Counterpart}
\end{figure}

As a result of astrometric calculations, a unique optical counterpart was determined for ULXs X-1, X-14, X-18, and X-24 while no optical source was detected at the corrected X-ray position of the ULX X-6. The remaining ULXs, X-3 and X-17, likely have multiple sources that exhibit unresolved or extended morphologies; therefore, they were not included in detailed analyses. A bright source was detected at the position of the ULX X-5, which matches the position of \textit{GAIADR3} source (source ID: 4676459695824016128) and 2MASS (J04173037-6247258) (one of the sources used for astrometric calculations) was not considered an optical counterpart. The corrected positions of ULXs and \textit{Chandra} source detection error ellipses are shown on the {\it HST} images, Figure \ref{F:Counterpart}, and photometric results are given Table \ref{T:Fotometri}. In the case of the {\it JWST} NIR counterparts, sources matching the optical counterparts of X-14 and X-24 were identified, which were also identified in the {\it HST/WFC3/IR} F160W images. Since the X-1 source was out of the field in both {\it JWST} and {\it HST} IR observations, it is unclear whether it has an IR counterpart. No detectable source was identified at the position of the remaining ULXs. NIR counterparts were not identified in the \textit{JWST} F770W and F2100W images due to the band feature and poor pixel resolution (0.111 pixels/arcsec); thus, these images were only used to assess the location of ULXs.

\begin{table*}
\centering
\caption{Vega magnitudes of the optical and infrared counterparts of four ULXs}
\begin{tabular}{cccccccccccccclll}
\hline
\multicolumn{2} {c} {} & Counterparts\\
Filters(date) & X-1 & X-14 & X-18 &X-24\\
\hline
\multicolumn{5}{c}{\textit{JWST/NIRCam}} \\
F277W(2023-06-30) & \textit{Out} & 21.45 $\pm$ 0.03 & \textit{Non} & 20.73 $\pm$ 0.02 \\
F150W(2023-06-30)& \textit{Out} & 20.83 $\pm$ 0.02 & \textit{Non} & \textit{Out} \\
F090W(2023-06-30) & \textit{Out} & 22.08 $\pm$ 0.03 & \textit{Non} & \textit{Out} \\
F090W(2023-09-14) & \textit{Out} & 22.05 $\pm$ 0.03 & \textit{Non} & \textit{Out} \\
F150W(2023-09-14)& \textit{Out} & 20.80 $\pm$ 0.01 & \textit{Non} & \textit{Out} \\
F277W(2023-09-14)& \textit{Out} & 21.44 $\pm$ 0.03 & \textit{Non} & 20.51 $\pm$ 0.02 \\
F300M(2023-09-14) & \textit{Out} & 20.39 $\pm$ 0.04 & \textit{Non} &19.61 $\pm$ 0.02 \\
F335M(2023-09-14) & \textit{Out} & 20.24 $\pm$ 0.05 & \textit{Non} & 18.52 $\pm$ 0.01 \\
\hline
\multicolumn{5}{c}{\textit{HST/UVIS1}} \\
F336W(2015-10-30) & 24.34 $\pm$ 0.08 & 22.53 $\pm$ 0.04 & 24.43 $\pm$ 0.09 & 24.07 $\pm$ 0.07 \\
F438W(2015-10-30) & 25.63 $\pm$ 0.06 & 22.64 $\pm$ 0.02 & 25.93 $\pm$ 0.12 & 24.97 $\pm$ 0.05 \\
F606W(2015-10-30) & 24.52 $\pm$ 0.03 &\textit{Out} & \textit{Out} & \textit{Out} \\
F814W(2017-09-12) &\textit{Out}& 22.12 $\pm$ 0.02 & 25.35 $\pm$ 0.13 & 23.42 $\pm$ 0.02 \\
F555W(2017-09-15) &\textit{Out}& 22.51 $\pm$ 0.01 & 25.26 $\pm$ 0.05 & 24.34 $\pm$ 0.03 \\
F814W(2017-09-26) &\textit{Out}& 22.11 $\pm$ 0.02 & 25.41 $\pm$ 0.12 & 23.39 $\pm$ 0.03 \\
F555W(2017-10-02) &\textit{Out}& 22.55 $\pm$ 0.02 & 25.27 $\pm$ 0.06 & 24.37 $\pm$ 0.03 \\
F814W(2017-11-07) &\textit{Out}& 22.20 $\pm$ 0.02 & 25.34 $\pm$ 0.12 & 23.43 $\pm$ 0.04 \\
F275W(2019-04-06) & 24.25 $\pm$ 0.05 & 22.09 $\pm$ 0.03 & 24.18 $\pm$ 0.06 & 24.43 $\pm$ 0.08 \\
\hline
\end{tabular}
\\ \textit{Out} indicates that the sources are outside the field of view, while \textit{Non} indicates that there is no counterpart \\
\label{T:Fotometri}
\end{table*}

Additionally, the F$_{X}$/F$_{opt}$ ratio is a useful diagnostic tool for identifying whether a source could be an AGN, based on the relative flux of its X-ray emission (F$_{X}$) compared to its optical emission (F$_{opt}$)  \citep{1982ApJ...253..504M,1991ApJS...76..813S,2010MNRAS.401.2531A}. Although there are no simultaneous optical and X-ray observations when considering the closest dates (within a 1-year interval), the high ratios for the counterparts of X-1, X-14, X-18, and X-24 exceed the value suggested for AGN or BL Lac types. Furthermore, as ULXs are found in crowded regions, it is likely that we detected false positive counterparts. To assess the number of false positive counterparts (N$_{FP}$), the false positive rates (FPR) were derived using a method similar to the study of \cite{2024MNRAS.527.2599A} (and reference therein). This approach involves calculating the probability of random matches within a specified positional error radius, 
allowing for the estimation of false positive identifications for a given set of counterparts. The average FPR is $\sim$ 5.7 percent for optical counterparts and $\sim$ 4.2 percent for NIR counterparts indicating that the counterparts determined in the error radius in this study are reliable.

\subsection{Variability of the counterparts}

\begin{figure}
 \resizebox{\hsize}{!}{\includegraphics{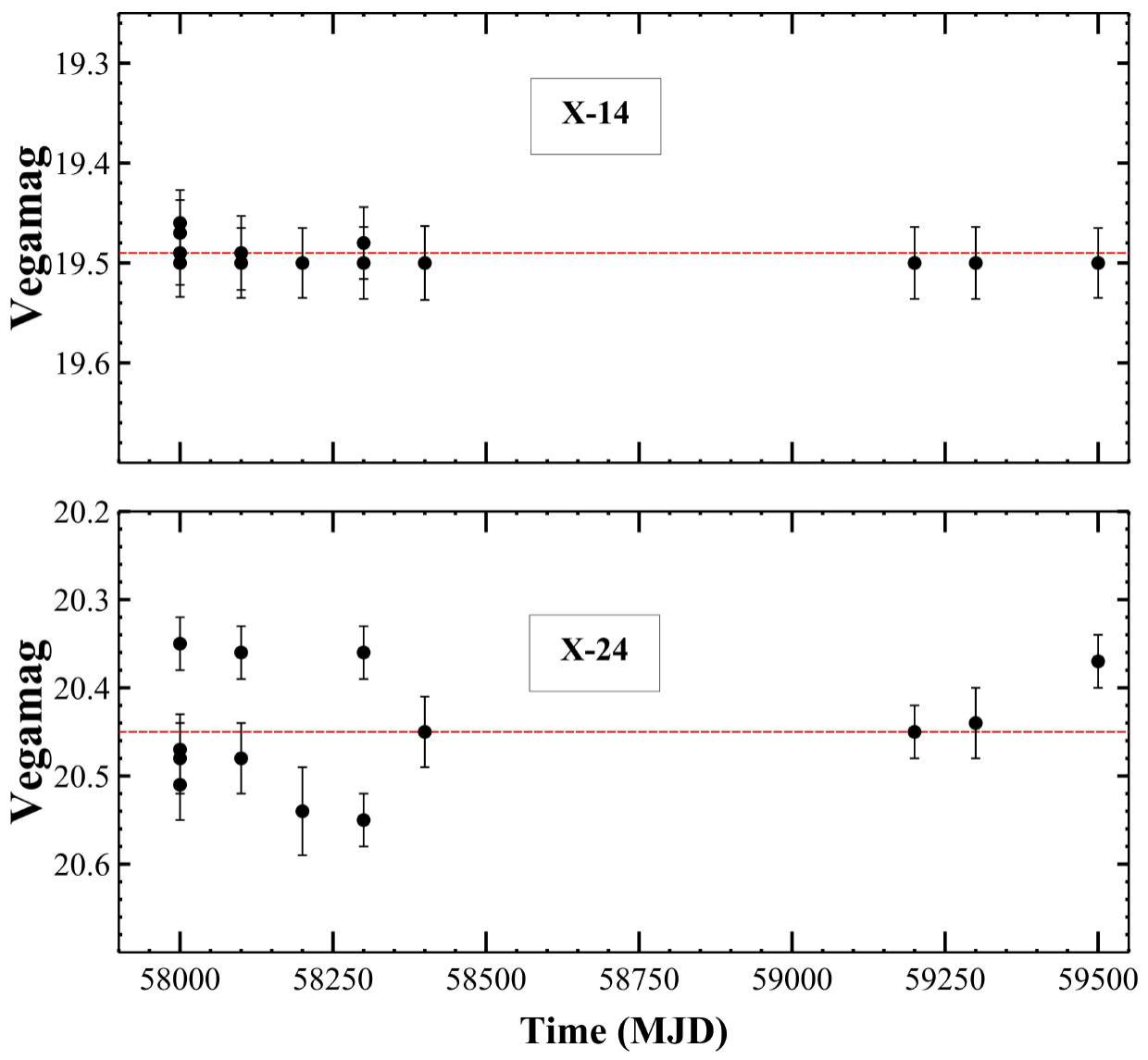}}
 \caption{Long-term light curves for X-14 (above) and X-24 (below) using 13 observations taken with the \textit{HST} F160W IR images. 
 The dashed red lines represent the average magnitudes derived from these observations.}
 \label{F:LC160}
\end{figure}

Given the relatively sufficient multi-epoch \textit{HST} and \textit{JWST} observations of the NGC 1559 galaxy (see Table \ref{T:obs}), we could place constraints on the variability of the observed emission from the counterparts. Therefore, through multi-wavelength observations of NGC 1559, we focused on the optical and IR variability of ULXs X-14, X-18, and X-24, which have unique optical counterparts. For ULX X-14, no variability was observed across \textit{HST} optical, IR, and \textit{JWST} NIR bands. No variation was observed in the optical emission of the X-18, and no sources were detected at the position of ULX X-18 in the \textit{HST} IR or \textit{JWST/NIRCam} images. The optical emission of the counterpart of X-24 remained nearly constant but notable ($\leq$ 0.2 mag) non-periodic variations were observed in the light curve of {\it HST} IR (F160W) and \textit{JWST} NIR (F277W) observations (Sept 2017 to Oct 2021), the long-term light curves for the X-14 and X-24 sources are displayed in Figure \ref{F:LC160}.

\subsection{SEDs and CMDs}

\begin{figure}
 \resizebox{\hsize}{!}{\includegraphics{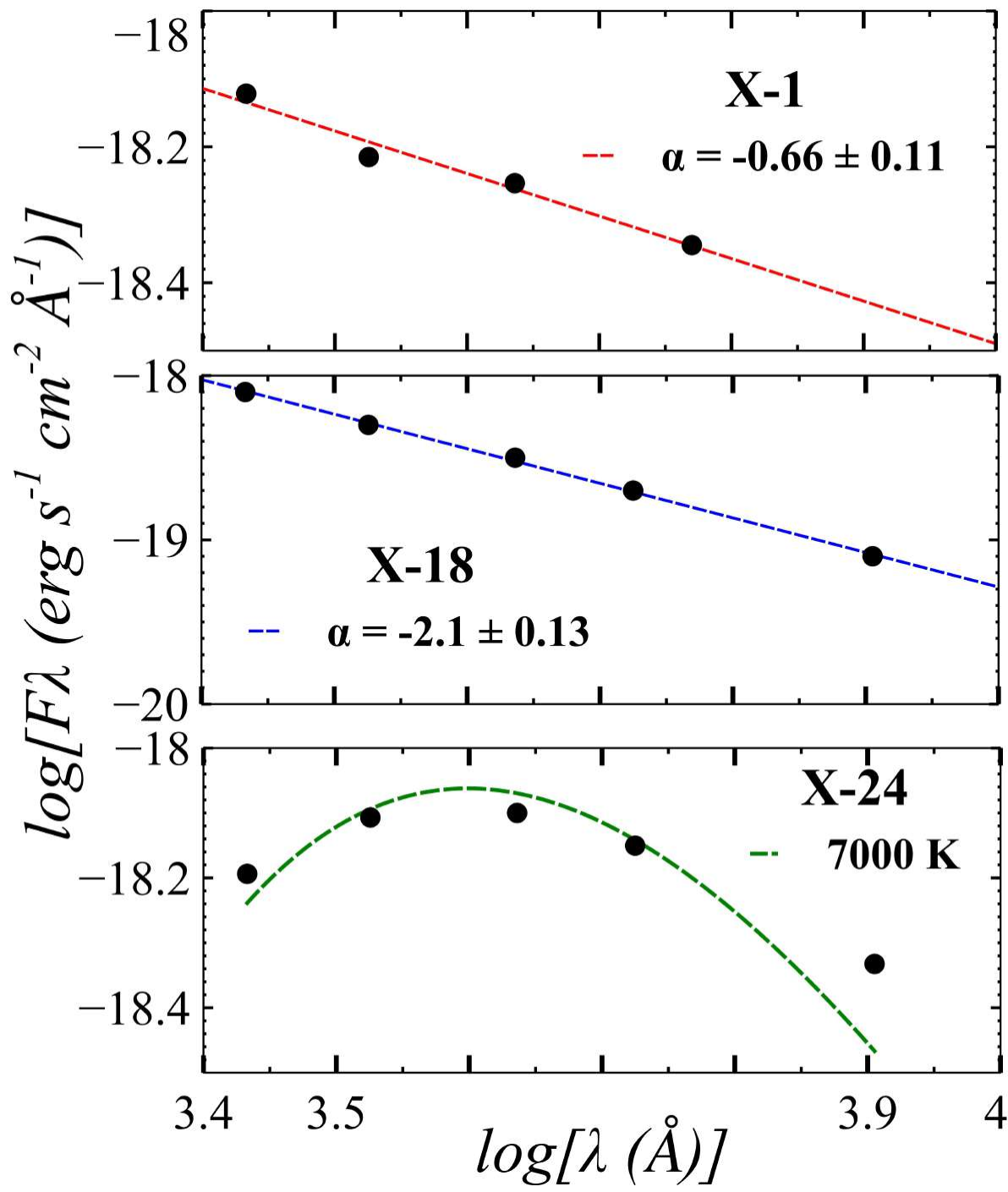}}
 \caption{SEDs of four counterparts of the ULXs X-1, X-18, and X-24 (from top to bottom). The {\it power-law} and black body models for ULXs X-1, X-18, and X-24 are shown by red, blue, and green dashed lines, respectively. All data are shown with filled black circles. The errors of data, taken as a systematic 0.05 mag, match the symbol size.}
 \label{F:SEDs}
\end{figure}

SEDs of the optical counterparts were constructed to obtain the spectral characteristics of the counterparts using flux derived from values given in Table \ref{T:Fotometri}. The wavelengths of the filters are selected as the pivot wavelength, obtained from {\scshape pysynphot}\footnote{https://pysynphot.readthedocs.io/en/latest/}. We attempt to constrain the nature of possible donor stars by fitting optical SEDs with a {\it blackbody} or a {\it power-law} (F $\propto$ $\lambda^{\alpha}$) models. Physically acceptable parameters were obtained only for X-1, X-18, and X-24. The SEDs of X-1 and X-18 were adequately fitted by \textit{power-law} models with spectral indices ($\alpha$) of -0.66 $\pm$ 0.11 and -2.1 $\pm$ 0.13, respectively, in the case of X-24, the SED is adequately represented by a \textit{blackbody} model with a temperature of 7000 K. The number of degrees of freedom (dof) for X-18 and X-24 is three and for X-1 is two. Optical SEDs for counterparts X-1, X-18, and X-24 are displayed in Figure \ref{F:SEDs}.

\begin{figure}
\resizebox{\hsize}{!}{\includegraphics{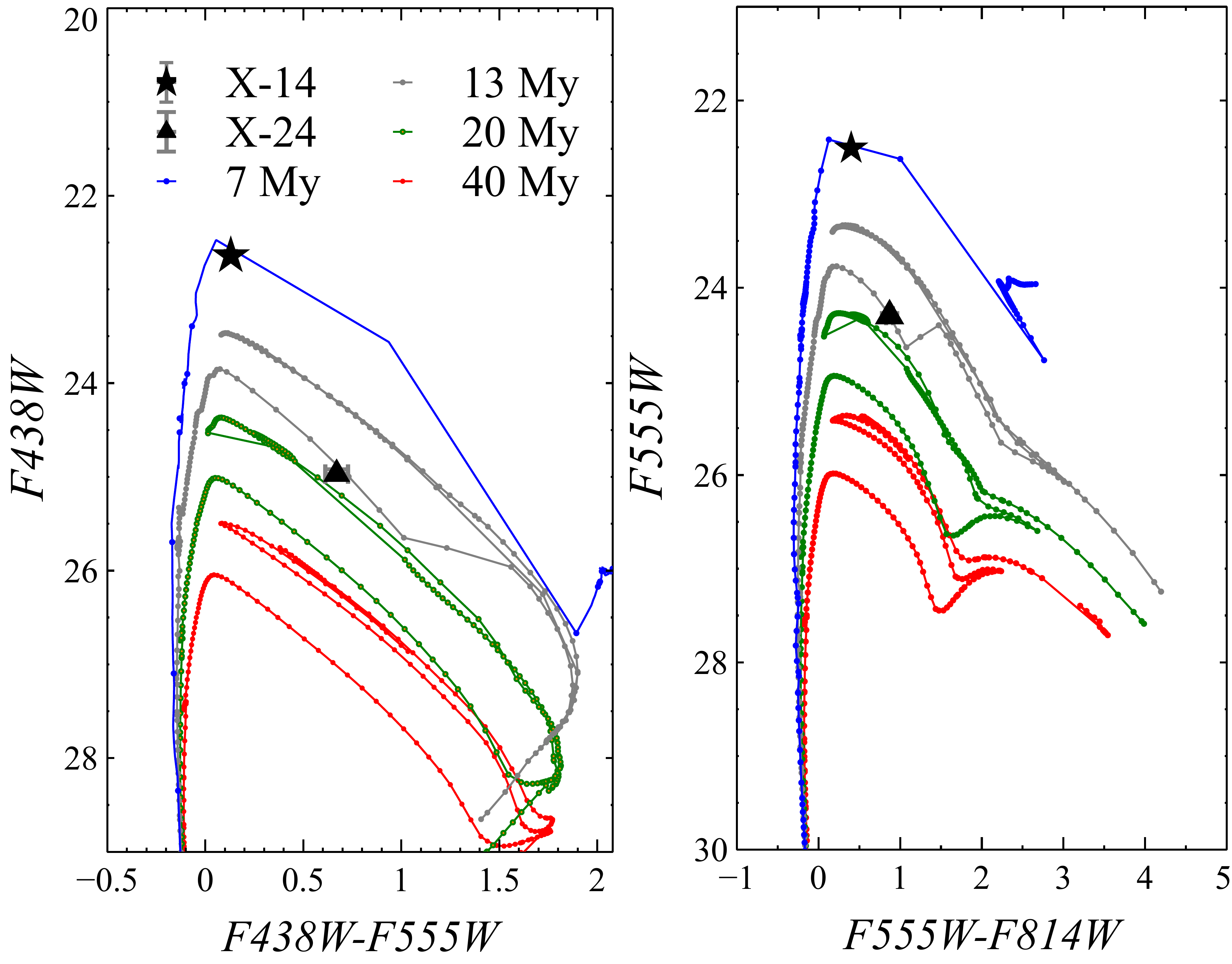}}
 \caption{ {\it HST/WFC3} color-magnitude diagrams for counterparts of X-14 (filled star) and X-24 (filled triangle). The isochrones were corrected for an extinction of A$_{V}$ = 0.04 mag.}
 \label{F:CMDs}
\end{figure}

Studying the stellar population around the ULX provides valuable insights into the characteristics of the donor star. Stars in the same region likely formed at similar times, allowing for more reliable estimates of the properties of donor candidates. Using CMDs of nearby stars, this approach helps to constrain the age and mass of the donor stars. Optical photometric and SED results suggested that their donor stars could dominate the optical emission for X-14 and X-24. Therefore, two CMDs were derived for the optical counterparts of X-14 and X-24: F438W (B) vs F438W$-$F555W (B-V), F555W (V) vs F555W$-$F814W (V-I) (see Figure \ref{F:CMDs}). Isochrones were obtained using solar metallicity of 0.02 and PARSEC models \citep{2012MNRAS.427..127B}. The distance modulus was calculated as 30.5 magnitudes, corresponding to the adopted distance of 12.6 Mpc.

\subsection{{\it Swift/XRT} data analysis of X-1}

\begin{figure}
 \resizebox{\hsize}{!}{\includegraphics{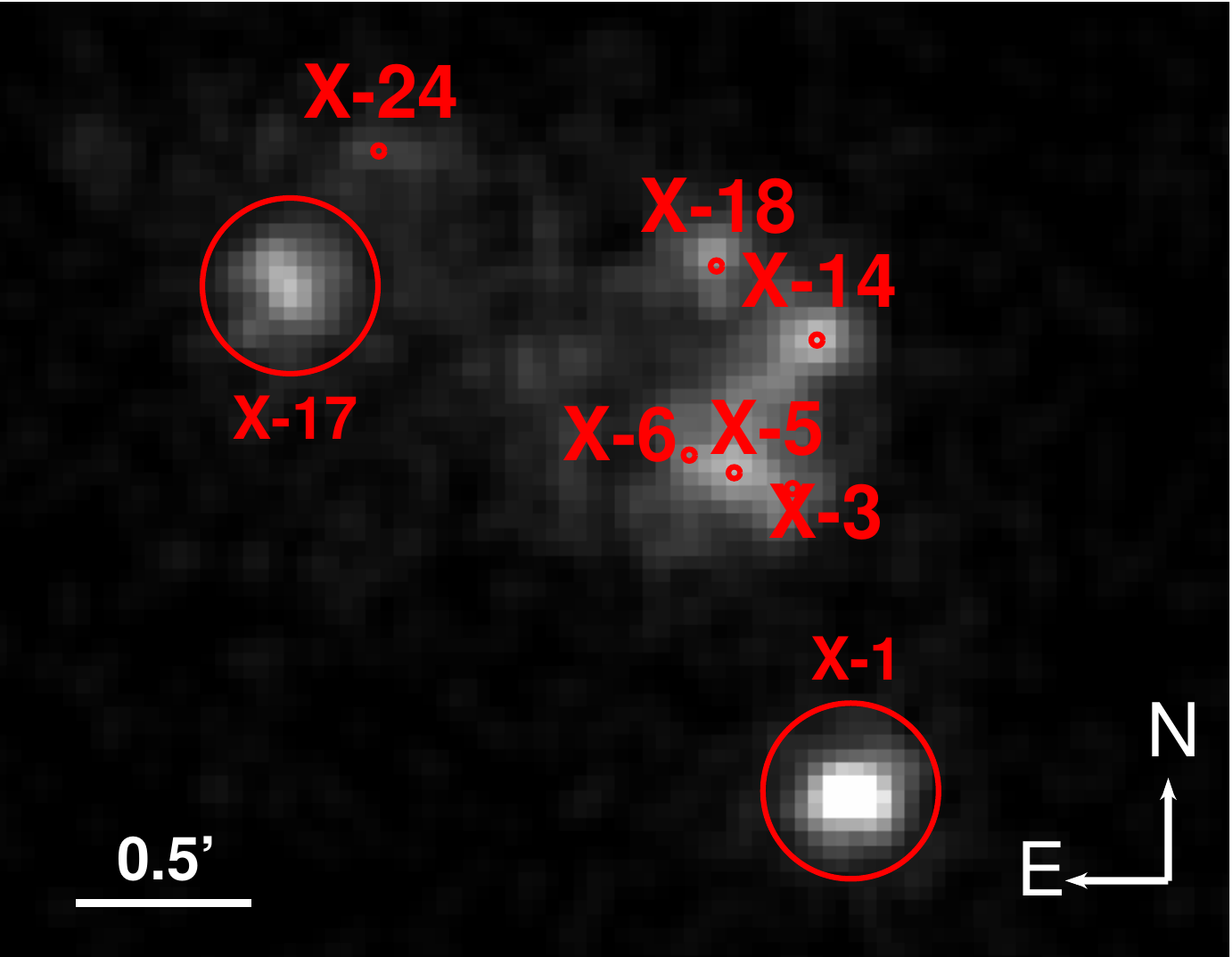}}
 \caption{The stack \textit{Swift/XRT} image of the galaxy NGC 1559. The image is smoothed with a 2 arcsec Gaussian. The ULXs are indicated with red circles on the image.}
 \label{F:XRT}
\end{figure}

The extensive set of \textit{Swift/XRT} observations allows us to conduct a detailed study of both the long-term variability and energy spectra of X-1, as it is the only isolated source without contamination among the eight ULXs.(see Figure \ref{F:XRT}). Although the study by \cite{2023ApJ...956...41M} mentioned the X-ray properties of X-1, we performed a more detailed analysis of this source, emphasizing its multi-wavelength characteristics and long-term X-ray evolution. By Utilizing 56 {\it Swift/XRT} Photon Counting mode (PC) observations, the temporal variability and spectral characteristics were examined. To obtain the source count rate, we used \textit{HEASoft} together with \textit{Swift/XRT}
data products generator\footnote{https://www.swift.ac.uk/user\_objects/index.php} \citep{2007A&A...469..379E,2009MNRAS.397.1177E}. For the long-term X-ray light curve and the time-averaged spectrum of X-1, circular extraction regions with radii of 15 arcseconds for the source and 45 arcseconds for the background were used. The source energy spectra were grouped using the FTOOLS {\it grppha} with at least 15 counts per energy bin. Spectral fits were performed on all available 0.3-10 keV spectral data to determine the best-fitting model for the time-averaged spectrum using {\scshape xspec} v12.13. A hydrogen column density (\textit{N$_{H}$}) component ({\it tbabs}) was included in all cases.

\begin{figure*}
\begin{center}
\includegraphics[angle=0,scale=0.45]{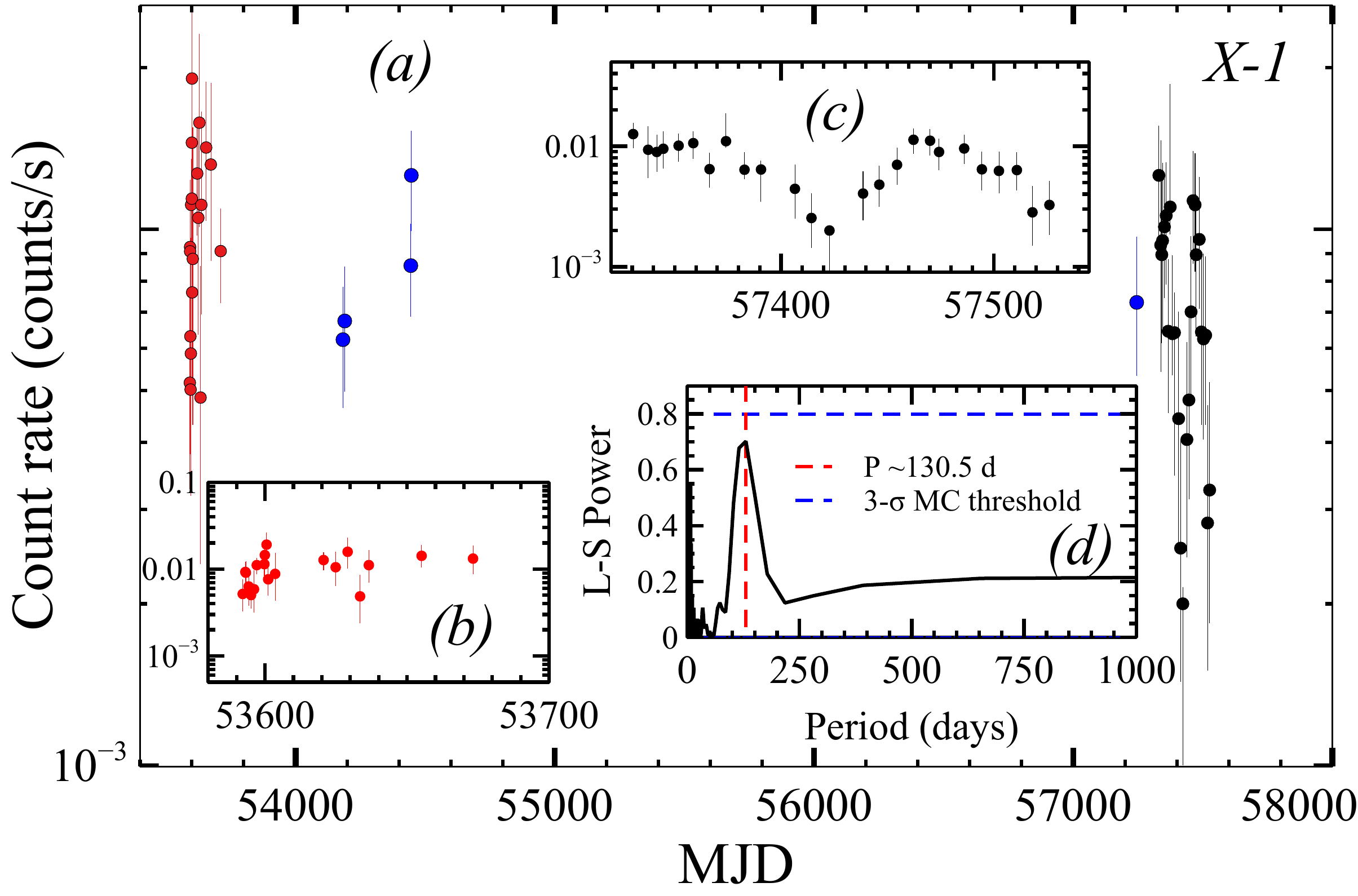}
\caption{Panel \textit{(a)}: Long-term S\textit{wift/XRT} light curve of the ULX X-1. The epoch-1 (\textit{panel b}), epoch-2, and epoch-3 (\textit{panel c}) observations are indicated by red, blue, and black-filled circles, respectively. Panel \textit{(d)}: Lomb-Scargle periodograms of the X-1. The periodic signal that peaks at 130.5 days (dashed red line) was found for epoch-3.}
\label{F:LCX1}
\end{center}
\end{figure*}

Using all \textit{Swift/XRT} observations, a long-term light curve of X-1 was constructed, displayed in Figure \ref{F:LCX1}. The ratio of maximum to minimum count rates for X-1 was determined to be one order of magnitude in the range of 0.3-10 keV. Since the \textit{Swift/XRT} observations are not continuous (e.g., observation dates, 2005, 2007, 2016), the light curve was divided into three epochs: 2005 (epoch-1), 2007 (epoch-2), and 2016 (epoch-3) to investigate possible periodic variations. As seen in panel \textit{C} of Figure \ref{F:LCX1}, a variation worth considering is noticeable only for epoch-3. Although all observations were used to investigate periodic variation, none was detected. Using the \textit{Lomb-Scargle periodogram, L-S} \citep{1976Ap&SS..39..447L, 1982ApJ...263..835S}, only in epoch-3, a low amplitude periodic modulation of 130.5 days was detected with false alarm probability (FAP) of $\sim$ 10$^{-4}$. To investigate whether this low-amplitude periodicity was due to random variability, following the procedures outlined in \cite{2022MNRAS.517.3495A}, a total of 2000 simulated red-noise time series were generated for each filter using Monte Carlo (MC) simulations, ensuring they had the same number of sampling points and parameter values. This approach provides a robust assessment of significance against potential false detections. To model the red-noise continuum, the L-S was applied to each simulated light curve, using a power-law index ranging from -1 to -2 and employing a multi-frequency approach. According to MC analysis, the identified period of 130.5 days exhibits a significance level close to 2-$\sigma$  (see Figure \ref{F:LCX1}), which makes the reliability of this period for the X-1 system questionable.

\begin{figure}
 \resizebox{\hsize}{!}{\includegraphics{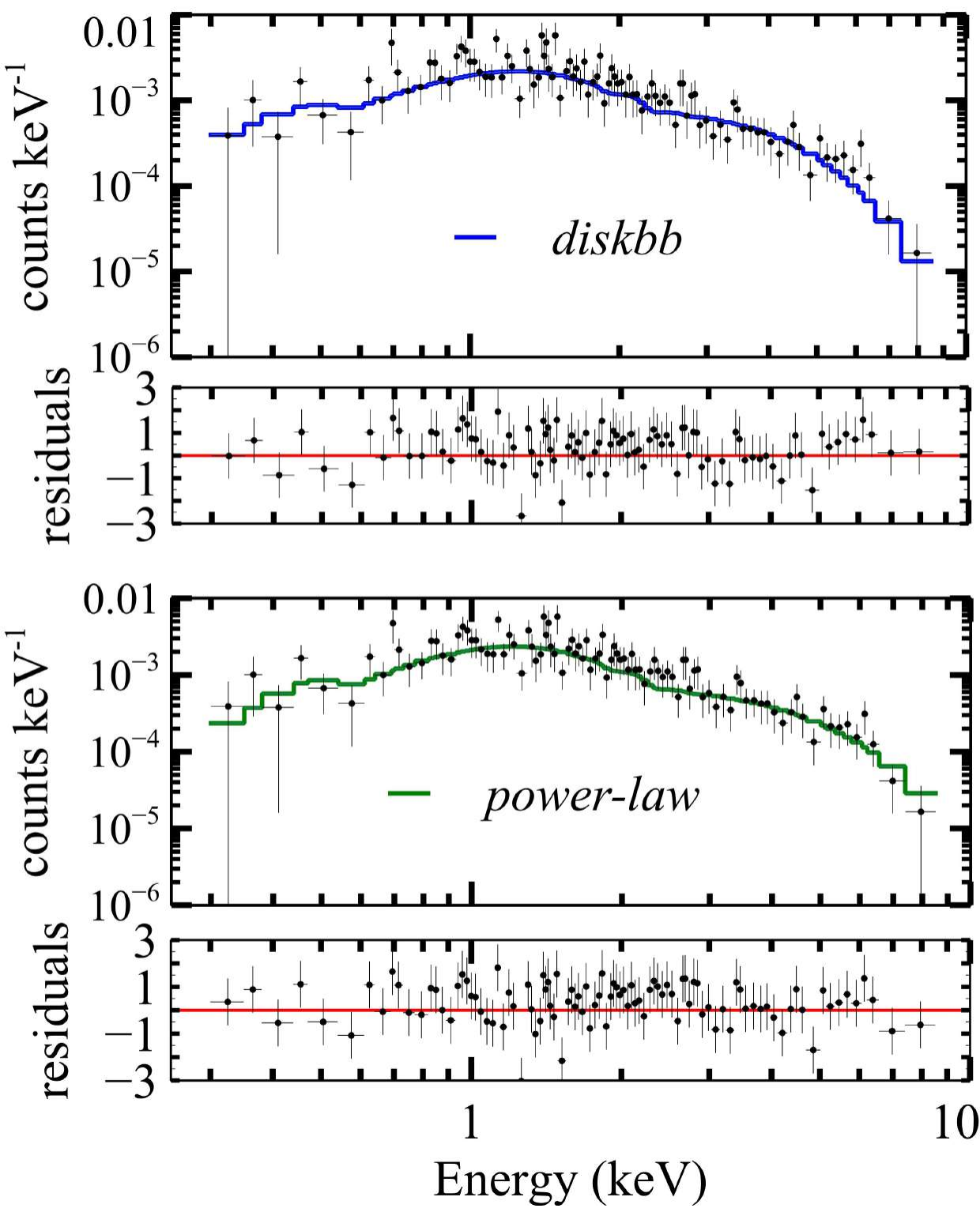}}
 \caption{\textit{Swift/XRT} time-averaged energy spectrum of the ULX X-1 with \textit{diskbb} (above) and \textit{power-law} (below) models.}
 \label{F: spec}
\end{figure}

The ULX X-1 is well-suited for applying a time-averaged spectrum due to its long-term variability and the low signal-to-noise ratio often present in individual observations. Consequently, the time-averaged spectrum for X-1 was obtained using data from 56 {\it Swift/XRT} observations and thoroughly analyzed to identify the well-fitted spectral models. The spectrum of X-1, along with the instrument response and ancillary files, was generated based on {\it Swift/XRT} observations. Due to the averaged source counts, the energy spectrum was grouped using \textit{FTOOLS grppha}, with at least 15 counts per energy bin. Spectral fits were then conducted over the 0.3-10 keV range to identify the best-fitting model for the time-averaged spectrum, utilizing \textit{XSPEC} v12.13. In all cases, a hydrogen column density (\textit{N$_{H}$}) component (\textit{tbabs}) was applied as part of the spectral modeling. Among the models tested, the \textit{power-law} and \textit{diskbb} models were found to be the best fits for the energy spectrum of X-1. The detailed parameters for these models are presented in Table \ref{T:spectra} and the time-averaged spectrum is displayed in Figure \ref{F: spec}.

\begin{table}
\caption{The {\it power-law} and {\it diskbb} model parameters of ULX X-1 from the {\it Swift/XRT} time-averaged spectrum}
\begin{tabular}{cccccll}
\hline
Parameters & Units & \textit{power-law} & \textit{diskbb} \\
\hline
N$_{\mathrm{H}}$ & 10$^{22}cm^{-2}$ & 0.20 $\pm$ 0.04 & 0.06 $\pm$ 0.03 \\
${\Gamma}$ & & 1.60 $\pm$ 0.12 & ... \\
N$_{\mathrm{{\it norm}}}$$^{\boldsymbol{a}}$ & 10$^{-3}$ & 6.60 $\pm$ 0.92 & 2.92 $\pm$ 0.78 \\
T$_{\mathrm{in}}$ & keV & ... & 1.69 $\pm$ 0.32 \\
L$_{\mathrm{X}}$$^{\boldsymbol{b}}$ & 10$^{39}$ ergs $s^{-1}$ & 8.87 $\pm$ 0.66 & 7.72 $\pm$ 0.64 \\
$\chi^{2}$/dof & & 81.57/96 & 80.68/96 \\
Null$^{\boldsymbol{c}}$ & & 0.85 & 0.87 \\
\hline
\label{T:spectra}
\end{tabular}
\\ Note: $^{\boldsymbol{a}}$ is the normalization parameter of the power-law and \textit{diskbb} ((r$_{in}$ km$^{-1}$)/(D/10 kpc)]$^{2}$ $\times cosi$)). $^{\boldsymbol{b}}$ is unabsorbed luminosity values calculated in the 0.3–10 keV energy band, adopted distance of 12.6 Mpc. $^{\boldsymbol{c}}$ is the Null hypothesis probability with 96 degrees of freedom.\\
\end{table}

\section{Results and discussion} \label{sec:4}

Following astrometric corrections, ULX counterparts of the eight previously identified ULXs were determined based on {\it HST} and {\it JWST} observations. Accordingly, a unique optical counterpart was identified for X-1, X-14, X-18, and X-24, while no source was detected for X-6, even at the 1-$\sigma$ confidence level. The presence of multiple optical and NIR sources within the astrometric error radius of both X-3 and X-17 makes it difficult to identify the most likely counterparts, so these sources were not discussed further. For X-5, a unique source was detected which is too extended to be an optical counterpart, and also it was matched with a \textit{GAIA} and \textit{2MASS} source (see Sec \ref{X-5}). On the other hand, X-1 is not covered by the {\it JWST} or \textit{HST} IR observation fields, and no IR counterpart(s) was detected within the error radius of the X-18. For the remaining ULXs, X-14 and X-24, unique infrared counterparts matching their optical counterparts were identified. The results obtained in this study and their discussion are given individually for each ULX as follows:

\subsection{ULX X-1}

ULX X-1 stands out as one of the sources among the eight identified ULXs that is isolated and free from contamination by nearby sources in all observations. The long-term light curve showed significant count rate variability (one order of magnitude) consistent with the results of \citep{2023ApJ...956...41M}. Additionally, low amplitude, low confidence ($\simeq$ 2-$\sigma$) potential periodic variability of 130.5-d was identified only in 2016 observations (epoch-3, see Figure \ref{F:LCX1}). The nature of this variability can be interpreted in the context of super-orbital periods observed in ULXs (e.g. \citealp{2019ApJ...873..115B,2020ApJ...895..127B}). Super-orbital periods in ULXs are mainly driven by precessing accretion disks, which can be affected by gravitational interactions with companion stars and varying mass transfer rates from those stars \citep{2023NewAR..9601672K}. The low FAP value of $\sim 10^{-4}$ indicates that this periodic variation is unlikely to result from random variability. However, the MC simulations evaluating this variation at the $3\sigma$ confidence level do not provide robust support for its classification as a real periodic variation. If the variation were random, it would align with the irregular and sometimes periodic fluctuations observed in AGNs \citep{2017MNRAS.464.3194B}. Additional observations are needed to confirm whether the periodic variation is physically meaningful.

The ULX X-1 in NGC 1559 displays spectral features that strongly suggest accretion onto a compact object, likely a stellar-mass BH or NS, with super-Eddington accretion processes playing a significant role. The \textit{power-law} model applied to the X-ray spectrum reveals an $\alpha$ = 1.60, indicating a hard X-ray spectrum typically associated with non-thermal emission, likely from high-energy processes occurring near the compact object. The parameters are also consistent with the values obtained by \cite{2023ApJ...956...41M}. 
Furthermore, the calculated unabsorbed L$_{X}$ $\sim$ 7.87 $\times$ 10$^{39}$ ergs s$^{-1}$ in the 0.3-10 keV energy band. Additionally, when modeled with a \textit{diskbb}, the inner disk temperature of 1.69 keV was found, pointing to a hot accretion disk, typically associated with very high accretion rates. This model also yields an L$_{X}$ $\sim$ 6.75 $\times$ 10$^{39}$ ergs s$^{-1}$ in the 0.3-10 keV energy range. The X-ray energy spectrum fits the \textit{diskbb} and \textit{power-law} models, with nearly identical fit statistics for both. Therefore, it is challenging to distinguish between the two models. On the other hand, a hard \textit{power-law} and a thermal disk component may suggest a complex accretion environment, where super-Eddington accretion may be accompanied by outflows or relativistic jets \citep{2001ApJ...552L.109K}. These findings suggest that this ULX is powered by a compact stellar object, where the quite high L$_{X}$ and spectral features are driven by super-Eddington accretion onto a BH or NS.

\begin{figure}
 \resizebox{\hsize}{!}{\includegraphics{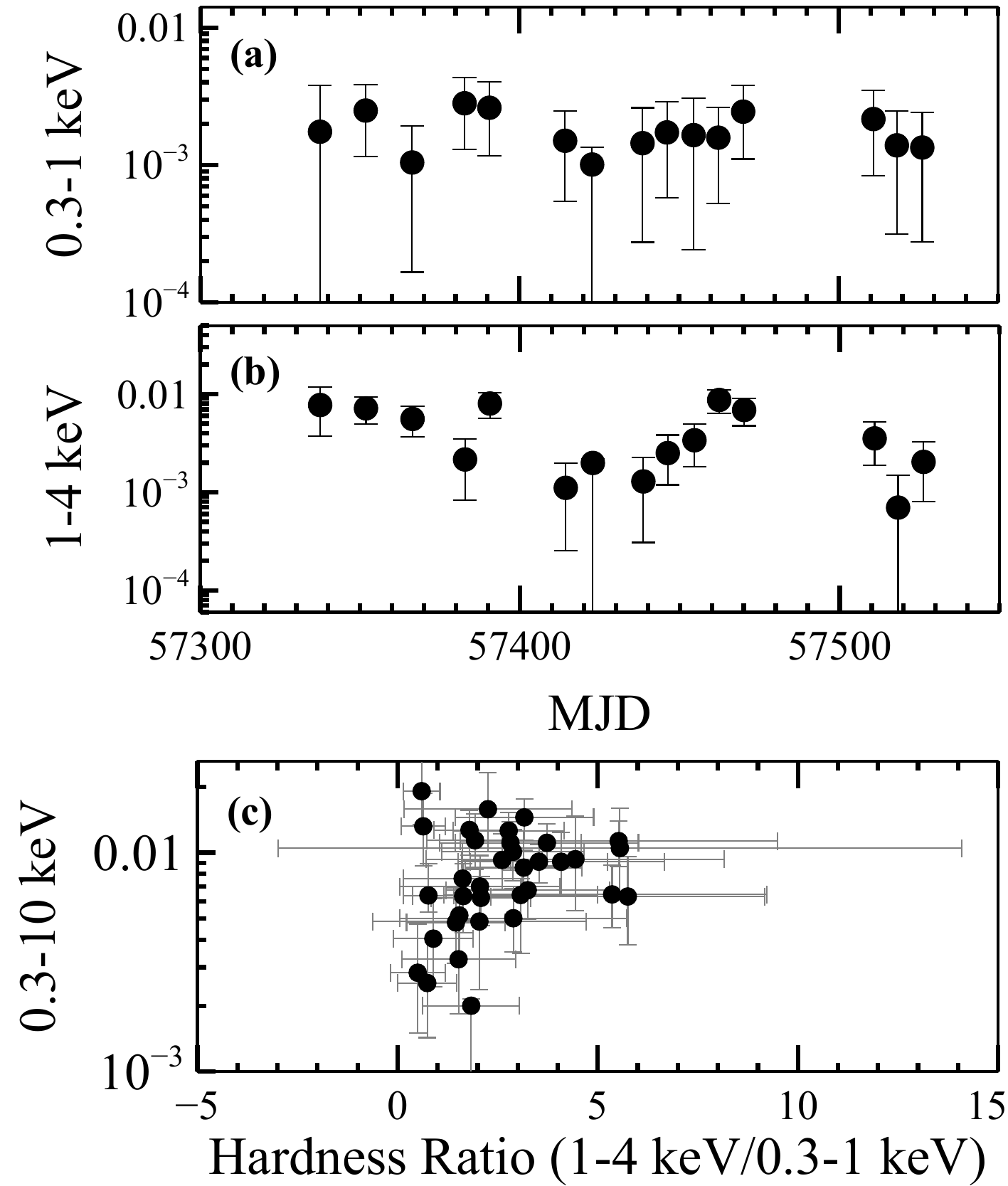}}
 \caption{Upper two panels show X-ray soft (0.3-1 keV) and hard (1-4 keV) count rates vs time diagrams, and the lower panel presents the hardness-intensity diagram of the ULX X-1.}
 \label{F:HR}
\end{figure}

To further investigate the long-term spectral evolution of ULX X-1, including a possible spectral state transition, temporal variations of the hard and soft count rates and a hardness-intensity diagram were plotted using \textit{Swift/XRT} observations (see Figure \ref{F:HR}). The hardness ratios (hard/soft) were calculated based on the hard and soft count rates in the energy ranges of 1–4 keV and 0.3–1 keV, respectively. The sinusoidal-like variation observed in epoch-3 for X-1 in the 0.3-10 keV energy range was not evident in the soft band while a similar variation was observed in the hard band. (see Figure \ref{F:HR}). The precession of the magnetic axis of the NS could also produce periodic modulations in the X-ray emission \citep{2003ASPC..302..241L,2003MNRAS.341.1020W,2020MNRAS.495L.139T}. Due to the insufficient availability of high-quality data, it remains uncertain whether X-1 undergoes spectral state transitions (see \textit{c panel} of the Figure \ref{F:HR}). If the source indeed displays spectral state transitions, the observed variability could originate from high-energy photons emitted from the innermost regions of the accretion disk \citep{2012MNRAS.420.1575K} or from the jet precession of the BH \citep{2018NatAs...2..443A,2023Natur.621..711C}. According to this scenario, the X-ray energy spectrum is more appropriately described by the \textit{diskbb} model (with T$_{\mathrm{in}}$=1.69 keV).

In the case of the optical counterpart of X-1, the optical SED was adequately fitted by a \textit{power-law} model with a spectral index of $\alpha$ $\sim$ -0.66. This suggests that the non-thermal emission from the accretion disk or jets is dominant, providing a clue that the accretion disk significantly contributes to the optical emission \citep{2011ApJ...737...81T,2022MNRAS.515.3632A}. We emphasize that the possible donor star of X-1 is quite faint in the UVIS1 bands (> 24.3 mag). We calculated its absolute magnitude in the F606W filter to be $\sim$ -6, which falls within the range of known optical candidates for ULXs (-3 < Mv < -8 ) \citep{2015NatPh..11..551F,2018ApJ...854..176V}. Although the nature of the donor star candidate cannot be further discussed due to the lack of infrared wavelength data, it could still have a sufficiently large mass to be detectable at this distance (12.6 Mpc), suggesting that the system may be a high-mass X-ray binary (HMXB) \citep{2020ApJ...890..150C}.

\subsection{ULX X-14}

The availability of data from different epochs in the same filters (e.g., \textit{JWST} F275W, and \textit{HST} F555W) for the counterparts enabled constraints on the optical and IR variability. Accordingly, it was determined that the optical emission of the X-14 was not significantly changed (in all cases $\leq $0.1 mag). Moreover, we found that the (N)IR counterpart matching the optical counterpart shows no variability in emission in both \textit{HST} IR and \textit{JWST/NIRCam} observations. This suggests that the donor star is the primary contributor to the optical and IR emissions. Moreover, the constant emission at the IR wavelength may indicate the absence of a variable jet. However, an SED could be created using optical and IR data, but a model with physically meaningful parameters could not be fitted. In addition, the CMDs plotted with the donor star candidate of X-14 and the surrounding stellar candidates indicate that the most likely age of the donor candidate is 7 Myr, with a mass of $\simeq$ 18 M$_\odot$. The donor candidate with an M$_{F555W}$ $\simeq$ -8 indicates that it is an HMXB at this distance of the NGC 1559.

\subsection{ULX X-18}

A possible donor star of X-18, similar to X-1, is faint in the optical bands but relatively bright in the UV band with calculated magnitudes of M$_{F555W}$ $\simeq$ -5 and M$_{F275W}$ $\simeq$ -6.
Furthermore, no counterpart was detected in the \textit{NIRCam} and IR bands within the astrometric error radius. The SED of X-18 is well represented by a \textit{power-law} model with an index of $\simeq$-2, which suggests that the accretion disk may dominate the observed optical emission.
The absence of a NIR counterpart and the interpretation of the SED could be attributed to various scenarios, such as the possibility that in ULXs, strong winds or super-Eddington accretion could disperse the cold material around the system, or that the optical emission likely originates from the hot accretion disk, making the IR counterpart undetectable due to disk dominance. If we consider the latter possibility, X-18 could be classified as an HMXB, potentially including a blue supergiant donor star.

\subsection{ULX X-24}

The candidate donor star of X-24 is faint in the UVIS1 bands with an M$_{F555W}$ $\simeq$ -5, but bright in the infrared, with an M$_{F335M}$ $\simeq$ -12 in the NIRCam.
The optical emission remained nearly constant (with variations of $\leq$ 0.1 mag in all cases), while a noticeable variation was detected in the IR emission (see Figure \ref{F:LC160}). This stability in the optical could suggest that it originates from a region possibly unaffected by the processes driving the variability in the IR and X-ray bands. The IR variability may be linked to mechanisms such as synchrotron emission from jets or thermal emission from surrounding dust, potentially influenced by periodic changes in the X-ray emission associated with the orbital dynamics 
\citep{2016ApJ...831...88D,2019ApJ...878...71L}.

\begin{figure}
\resizebox{\hsize}{!}{\includegraphics{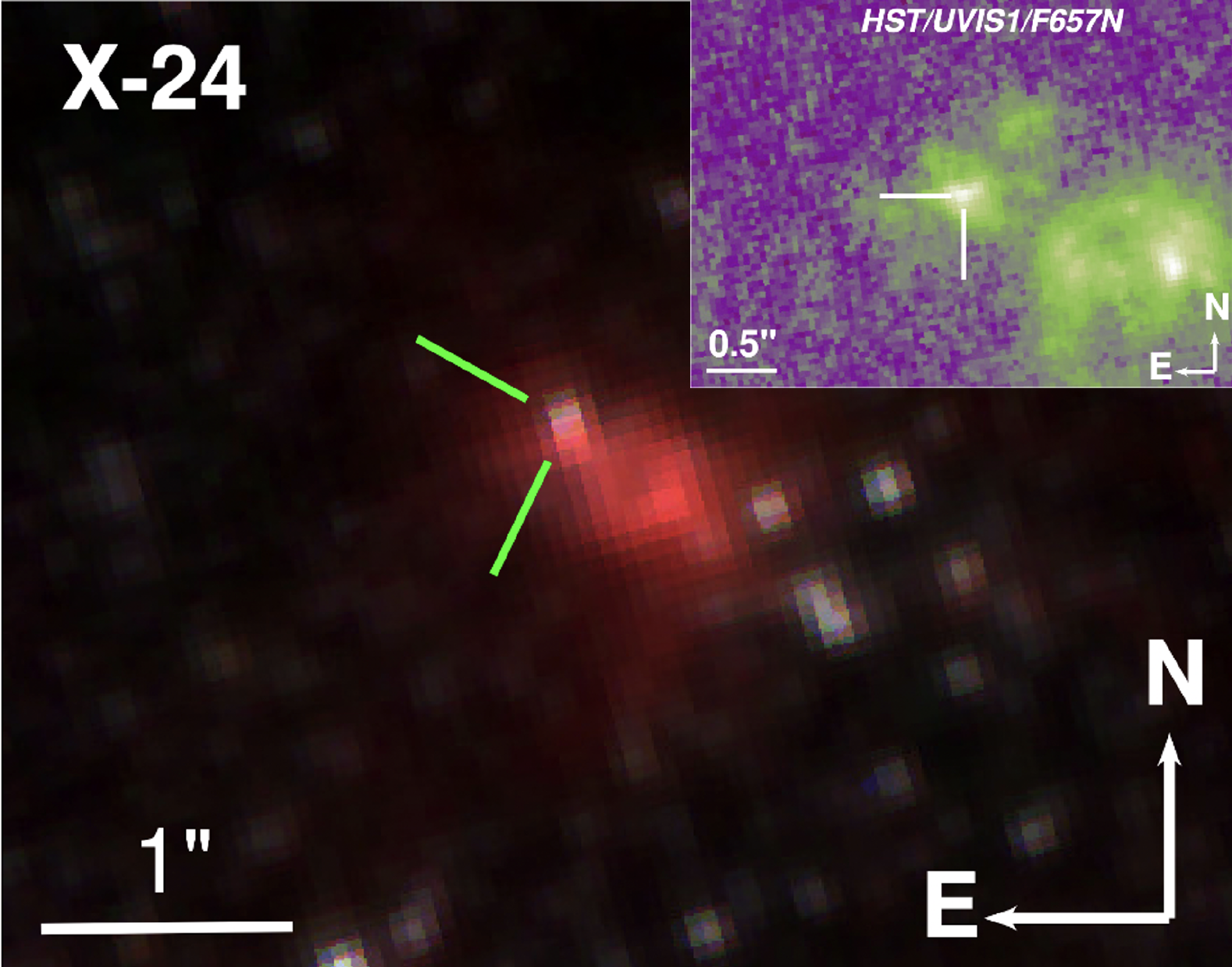}}
\caption{JWST/NIRCam image of the region around ULX X-24, in the broad-band filters F335M (red), F300M (green), and F277W (blue). The position of X-4 is also shown on the \textit{HST} UVIS F657N image in the upper right corner. White and green bars indicate the position of the donor star candidate of X-24.}
\label{F:env24}
\end{figure}

\begin{figure}
 \resizebox{\hsize}{!}{\includegraphics{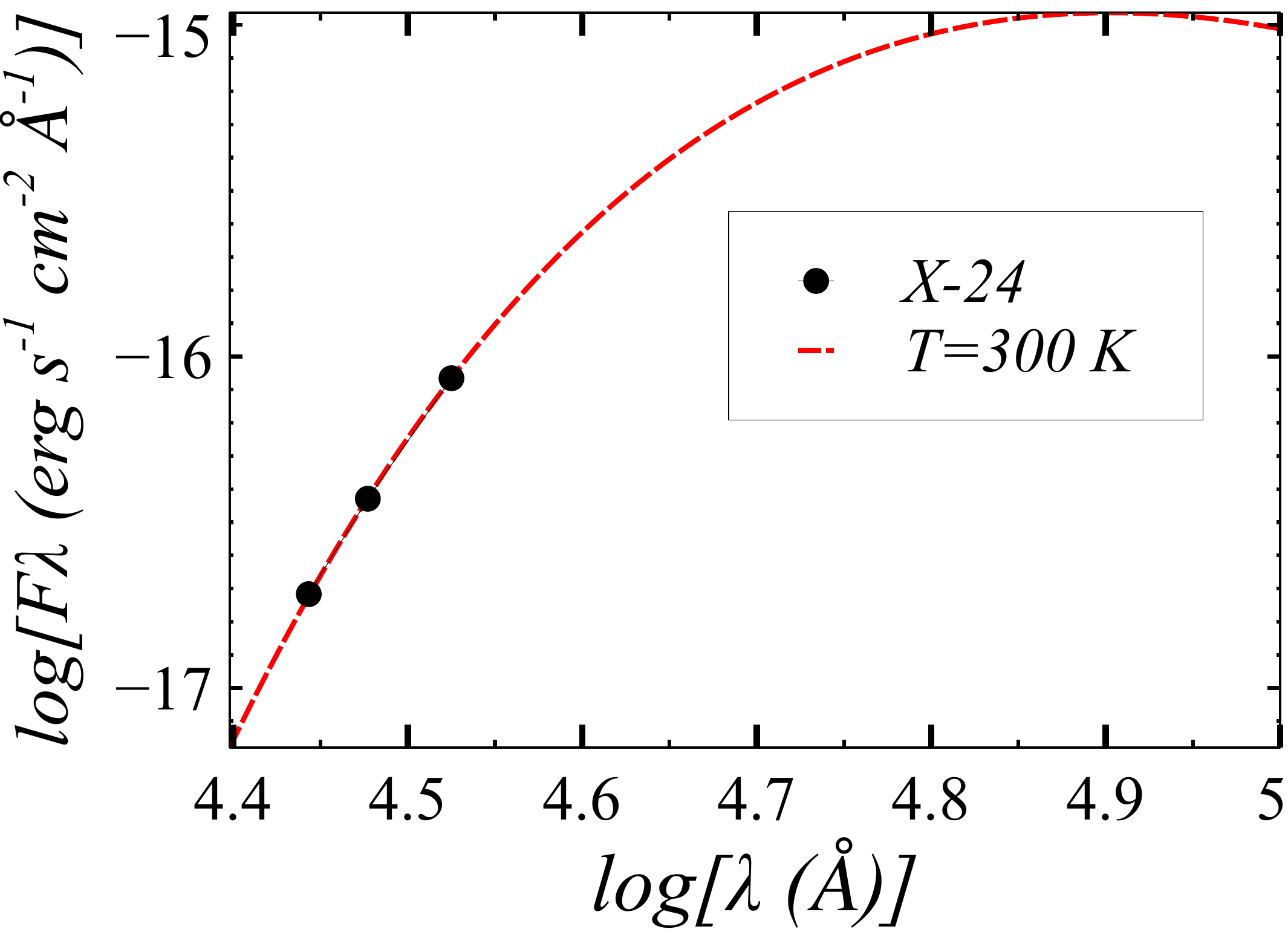}}
 \caption{The NIR SED of the donor star candidate for X-24. The dashed red line indicates a \textit{blackbody} temperature of 300 K. The errors, taken as a systematic 0.05 mag, match the symbol size.}
 \label{F:buble24}
\end{figure}

The optical SED of X-24 is modeled as a \textit{blackbody} with a temperature of 7000 K, suggesting that the donor star may be the source of the optical emission. Moreover, as seen in Figure \ref{F:env24}, X-24 is located in a dense region, likely composed of dust and gas, in the NIRCam image, which could be the source of the NIR emission observed from the position of donor star candidate. The \textit{HST} H$_{\alpha}$ (F657N) image also reveals the presence of an H II region at the same location (see Figure \ref{F:env24}). Additionally, the excess observed in the F814W image (see Figure \ref{F:SEDs}) may suggest that the system is surrounded by gas and dust, or potentially a circumbinary disk (\cite{2023MNRAS.526.5765A}, and references therein). Due to the variable NIR emission, only the SED of three simultaneous filters (F277W, F300M, and F335M) fits a \textit{blackbody} model with a temperature of 300 K (see Figure \ref{F:buble24}). This temperature is not high enough to come from the donor star. Therefore, it is likely that the observed NIR emission is due to gas and dust in the vicinity of the system, while the variation in the NIR emission could be attributed to the presence of a variable jet (\cite{2019ApJ...878...71L} and references therein).

According to the results obtained from the CMDs, the possible age of X-24 was found as 12 Myr, and according to this age, the mass of the donor candidate was derived as 12 M$_\odot$. Considering the optical absolute magnitudes as well as the mass of the donor star candidate, the system is most likely a supergiant in HMXB \citep{1981Ap&SS..80..353S, 1982BICDS..23....2S}. Although the donor candidate could appear as a red supergiant (RSG) based on its colors, the temperature obtained from the optical SED could be considered quite high for an RSG \citep{2013ApJ...767....3D} but relatively low for an HMXB supergiant (e.g. OB type). Furthermore, this relatively low temperature (7000 K) could be considered an additional indication of gas and dust in the surroundings, which may obscure the optical emission and result in a lower observed temperature.

In a previous study \citep{2023ApJ...956...41M} a 7500 s periodic modulation from X-ray data has been interpreted as an orbital period for X-24. The authors suggest that this observation is consistent with a scenario involving Roche-lobe overflow in a binary star system. In this context, material from the donor star is transferred to the compact object, likely a BH, leading to strong X-ray emissions. They have claimed that the donor star of X-24 could be an LMXB which is losing mass through Roche-lobe overflow. Furthermore, the periodic behavior and high X-ray luminosity indicate that X-24 may be a stellar-mass BH within a compact binary system. However, in contrast to them, the multi-wavelength properties indicated that the possible donor of X-24 could be HMXB. Given that the ULX X-24 is an HMXB supergiant, regardless of the mass of a compact object, the 7500 s orbital period presents a physically unlikely scenario for Roche lobe overflow. For instance, For a compact object mass of 3 M$_\odot$, the Roche lobe radius-to-separation ratio ({\it R$_{L}$/a}), calculated using the \textit{Eggleton} equation \citep{1983ApJ...268..368E}, is approximately 0.5, whereas for 80 M$_\odot$, it drops to about 0.2. With the fixed parameters of donor mass of 12 M$_\odot$ and a 7500 s orbital period, the separation remains too small for the donor to fill its Roche lobe. Thus, the periodic variation is more likely a quasi-periodic oscillation (QPO), spin period, or random variability, rather than an orbital period.

\section{Summary and Conclusions} \label{sec:5}

In this study, we conducted an X-ray and multiwavelength analysis of eight ULXs in NGC 1559. We summarize our key findings as follows:

\begin{enumerate}

\item Using astrometric calculations, we identified unique optical counterparts for ULXs X-1, X-14, X-18, and X-24. For IR observations with \textit{JWST}, we found only two ULXs (X-14 and X-24) that matched their optical counterparts. For the remaining ULXs, X-3 and X-17, multiple counterparts were found for both optical and IR images. Also, a source was matched with the position of the X-5 from the GAIA and 2MASS catalogs, while no counterpart(s) were found for X-6 in either \textit{HST} or \textit{JWST}. 
 
\item The models of the X-ray spectrum of X-1 did not allow us to determine whether its compact object is a BH or a NS. The long-term X-ray light curve of X-1 shows a count rate variation of almost one order of magnitude. Additionally, the optical counterpart of X-1 was detected as very faint in the UVIS1 bands. The constructed SED suggests that the donor star candidate is not the dominant source of the observed optical emission.

\item No variation was observed in the X-14 emission from both optical and NIR counterparts. The main source of the observed emission is likely the donor star candidate. CMDs indicate that this donor candidate could be a component of the HMXB system

\item Although the stable optical emission of X-18 indicates a significant contribution from the donor star, its SED suggests that non-thermal processes from the accretion disk could also contribute to the optical emission. No NIR counterpart was detected for X-18, possibly due to surrounding dust or gas being dispersed by strong winds or super-Eddington disk accretion processes.

\item For the donor candidate of X-24, the constant optical emission, along with the source SED that is well-fitted by a \textit{blackbody} at 7000 K, suggests that the observed emission may originate from the donor star. Additionally, the variable IR emission detected for this ULX may indicate the influence of a variable jet effect.

\item In the study by \cite{2023ApJ...956...41M}, X-24 was reported to have an orbital period of 7500 s based on X-ray analysis. However, our analysis of CMDs suggests that X-24 has a supergiant donor star. Therefore, we interpreted that the reported period as being too short to represent the orbital period of an HMXB system.

\end{enumerate}

\begin{acknowledgements}
This paper was supported by the Scientific and Technological Research Council of Türkiye (TÜBİTAK) through project number 122C183. AA acknowledges support provided by the TÜBİTAK through project number 124F004. We would like to thank the reviewer for the thoughtful comments and efforts towards improving our manuscript.
\end{acknowledgements}

%

\bibpunct{(}{)}{;}{a}{}{,}
\bibliographystyle{aa}
\bibliography{ngc1559}

%


\end{document}